\documentclass[aps,prl,twocolumn,showpacs,superscriptaddress,amssymb,floatfix,tightenlines]{revtex4}

\newcommand{\beq}{\begin{equation}}
\newcommand{\eneq}{\end{equation}}
\usepackage{graphicx}
\usepackage{amsmath}
\usepackage{epsfig}
\usepackage{dcolumn}
\usepackage{bm}
\usepackage{theorem}
\usepackage{amsmath,amssymb}

\newtheorem{theorem}{Theorem}[section]

\newenvironment{proof}[1][Proof]{\begin{trivlist}
\item[\hskip \labelsep {\bfseries #1}]}{\end{trivlist}}

\newcommand{\qed}{\nobreak \ifvmode \relax \else
\ifdim\lastskip<1.5em \hskip-\lastskip \hskip1.5em plus0em
minus0.5em \fi \nobreak \vrule height0.75em width0.5em
depth0.25em\fi}

\input{epsf}

\newcommand{\ket}[1]{\left| #1 \right\rangle}

\newcommand{\Tr}{\mathrm{Tr}}

\begin{document}

\title{A pairwise additive strategy for quantifying  multipartite entanglement}
\author {Gerardo~A. Paz-Silva}
\altaffiliation{E-mail: {\tt gerapaz@univalle.edu.co}}
\affiliation{Departamento de F\'isica, Universidad del Valle, A.A.
25360, Cali, Colombia}
\author {John H. Reina}
\altaffiliation{E-mail: {\tt j.reina-estupinan@physics.ox.ac.uk}}
\affiliation{Departamento de F\'isica, Universidad del Valle, A.A. 25360, Cali, Colombia}
\affiliation{Institut f\"ur Theoretische Physik, Technische Universit\"at Berlin, Hardenbergstr. 36, 10623 Berlin, Germany}

\date{\today}

\begin{abstract}

Based on the idea of measuring the factorizability of a given
density matrix, we propose a pairwise analysis strategy for
quantifying and understanding multipartite entanglement. The
methodology proves very effective as it immediately guarantees,  in
addition to the usual entanglement properties,  additivity and
strong super additivity. We give a specific set of quantities that
fulfill the protocol and which, according to our numerical
calculations, make the entanglement measure an LOCC non-increasing
function.  The strategy allows a redefinition of the structural
concept of global entanglement.
\end{abstract}


\maketitle

In a previous work of ours we presented a
geometrical scenario to quantify multipartite entanglement~\cite{PazReina2006-1}. To account for
the known problems exhibited by the Meyer-Wallach
measure of entanglement~\cite{Meyer-Wallach}, we proposed a quantity which can be evaluated for any bipartition of a
multipartite quantum state. An average of such a quantity over all
bipartitions would then give a measure of the entanglement of the
state independently  of the considered bipartition. However, the
quantity proposed in Ref.~\cite{PazReina2006-1}  had the
inconvenience that it did not exhibit local unitary invariance for
transformations on the first qubit.

To overcome this, in this work we choose an approach whereby,
instead of taking all of the possible state bipartitions, we
consider all of the possible pairs. This introduces the need for a
quantity that is capable of characterizing the degree of
entanglement between any two given qubits. Thus, we introduce a
probe quantity $\mathcal{P}(A,B)= \mathcal{P} (\rho_{A B})$ which
measures the degree of non-factorizability between two qubits, not
in the sense of the tangle $\tau$~\cite{Wooters-tangle}, but with
the idea of measuring the factorizability of the density matrix,
namely how feasible it is to write the density matrix as $\rho_{AB}
= \rho_A \otimes \rho_B$. Note that this idea is closely related to
the concept of entanglement in a pure two-qudit
scenario~\cite{Wooters-tangle}.

Many measures of multipartite entanglement have been
proposed through different mechanisms~\cite{Vedral-entropy,Coffman-3
tangle,Vidal-monotones,Milburn-exchange,Acin-decomposition}. In
particular several authors~\cite{Emary,Oliveira-GEM,Scott,PazReina2006-1}
have tried to describe the entanglement in a pure multipartite state
through quantifying the available entanglement in a specific
bipartition of the state and then taking on all bipartitions. This
approach involves some  issues: first of all, one must be able to
find an appropriate unitary invariant quantity which does the trick,
and second, for large numbers of qudits the amount of possible
bipartitions is too big.
Instead, we  consider
a more economic approach by resorting to all the possible two-qudit
density matrices~\footnote{One can see that by
calculating the number of nonequivalent bipartitions and the number
of all possible pair of qubits, after $N=5$, the number of
bipartitions outnumbers the number of pairs.}, expand on the consequences
and advantages of this strategy, and propose specific candidates for
the $\mathcal{P}$ measure.

We define our entanglement measure $\mathcal{M}$ as an arithmetical
average of the quantity $\mathcal{P}$
 \beq \mathcal{M}_{\mathcal{P}} =
\mathcal{N} (\mathcal{P}(A,B))\left(C^N_2\right)^{-1} \sum_{(A,B)}
\mathcal{P} (A,B) \ ,
 \eneq
where $C^N_2 \equiv \left(\begin{array}{c}
                N \\
                2
              \end{array}\right)$,
and the sum is over all possible non-equivalent arrays of pairs of
qubits. Note the dependence of the normalization factor $\mathcal{N}
(\mathcal{P}(A,B))$ on the quantity we use as a probe: $\mathcal{N}
(\mathcal{Q}_C(A,B)) = 1$ and $\mathcal{N} (\mathcal{F}r(A,B)) =
2-\delta_{N,2} $ (see below for definitions of $\mathcal{Q}_C$ and $\mathcal{F}r$).  The  case of mixed states would require the
extension \beq \mathcal{M}(\rho) = \min \sum p_i \mathcal{M}(\rho_i)
\ , \eneq where the minimum is intended over all possible
decompositions. Alternatively we can rewrite the minimization
condition as follows. Let $\rho$ be a generic (mixed or pure)
$n$-partite density matrix, and let $\rho_{s_i s_j}$ be the reduced
density matrix for qubits $s_i$ and $s_j$. In terms of the probe
quantities, the minimization condition reads
\begin{eqnarray}\nonumber
\mathcal{M}(\rho) &=& \min \sum p_l  \mathcal{M}(\rho_l)
= \min \sum_l p_l \Big(\sum_{s_i \neq s_j} \mathcal{P}(\rho^{(l)}_{s_i s_j}\Big)\\
&=& \min \sum_{s_i\neq s_j} \sum_l p_l\mathcal{P}(\rho^{(l)}_{s_i
s_j}) \ .
\end{eqnarray}
This implies that the minimization condition is equivalent to
minimizing the value of $\mathcal{P}$ for each two qubit reduced
density matrix, that is $ \mathcal{P}(\rho_{s_i s_j}) = \min \sum_l
q_l \mathcal{P}(\rho^{(l)}_{s_i s_j})$, where $\rho^{(l)}_{s_i s_j}$
may be mixed, with three simultaneous constraints: i) All reduced
density matrices must have the same coefficients, if the two qubit
reduced density matrices are minimized by $\rho_{s_i s_j} = \sum_l
q^{(s_i s_j)}_l \rho_{s_i s_j}^{(l)}$ then $q^{(s_i s_j)}_l = f_l$
for all pairs $(s_i s_j)$, ii) The set of two qubit reduced density
matrices $\rho_{s_i s_j}^{(l)}$ correspond to an $n$-partite pure
density matrix $\rho_l$ for each $l$, and iii) $\rho$ is expanded by
$\sum p_l \rho_l$. This definition is consistent when $\rho$ is a
pure density matrix: condition iii) requires that there is only one
non vanishing $q_l$, thus automatically guaranteeing conditions i)
and ii) and reducing to our previously defined measure. This
alternative form of the minimization condition,  a {\it pairwise
minimization condition}, will be used in the proof of  the
properties described below. In this way we gain some properties that
are desirable for entanglement measures, but that may not be so
conventional or easy to satisfy~\cite{Christandl}\footnote{In this
section we will explore mainly two properties: additivity and strong
super additivity. As we want it to be consistent with the
normalization condition $\mathcal{E}(EPR) = \mathcal{E} (GHZ_N)$ we
are not casting them in the traditional way. The reader must not be
deceived however, we will also show in subsequent sections that
replacing the above condition for a suitable normalization condition
where $\mathcal{E}(EPR) \leq \mathcal{E} (GHZ_N)$, and thus removing
the normalization factors on $\mathcal{M}$, leaves us with
traditional additivity and strong super additivity, thus showing the
nice properties of our strategy.}. Additivity is one good example:
given two pure density matrices, say $\sigma$ and $\eta$, and an
entanglement measure $\mathcal{E}$, then \beq \label{add}
\mathcal{E}(\sigma \otimes \eta) \geq \mathcal{E}(\sigma\otimes 0_E)
+ \mathcal{E}(0_E \otimes \eta) \ , \eneq where $0_E$ is a generic
separable density matrix. If for all $\sigma$ and $\eta$, equality
holds then we say the measure is {\it fully additive}; if only the
inequality is true the measure is {\it sub additive}. We make
explicit the inclusion of $\otimes 0_E$ and $0_E \otimes$ to avoid a
situation whereby, if $\sigma$ and $\eta$ are EPR density matrices,
then we would have that $\mathcal{E}(\sigma) + \mathcal{E}(\eta) = 2
= \mathcal{E}(\sigma \otimes \eta)$, which is unsatisfactory with
$\mathcal{E}(\ket{EPR}) = \mathcal{E}(\ket{GHZ_N})$.

For the sake of clarity, consider,
for example, the case of two two-qubits density matrices
$\sigma_{1,2}$ and $\sigma_{3,4}$. Then $
\mathcal{E}(\sigma_{1,2}\otimes \sigma_{3,4}) = (\mathcal{P}(1,2) +
\mathcal{P}(3,4))/6 $, since
$\mathcal{P}(1,3)=\mathcal{P}(1,4)=\mathcal{P}(2,3)=\mathcal{P}(2,4)=0$.
Then additivity is fulfilled since  $\mathcal{E}(\sigma_{1,2}
\otimes 0_E) = \mathcal{P}(1,2)/6$, and $\mathcal{E}(0_E \otimes
\sigma_{3,4}) = \mathcal{P}(3,4)/6$. A similar argument holds for
arbitrary  $N$-qubit pure density matrices. The general proof that
$\mathcal{M}$ is additive requires the following extra condition.
Let
$\rho = \sum p_i \rho_i$, and $\eta = \sum n_i \eta_i$, be two
generic density matrices in their $\mathcal{M}$-minimizing
decompositions of the qubits $\{ i_p\}$, and $\{i_n\}$ respectively,
with $N_s = {\rm dim}\{i_s\}$. Then the corresponding
minimizing decomposition  $\rho\otimes\eta = \sum p_i
\rho_i\otimes\sum n_j \eta_j$. We prove this by construction, here
$\rho$ and $\sigma$ are two qubit density matrices.  Let us assume that the minimizing
decomposition of $\rho\otimes\sigma$ is given by
$\rho_{AB}\otimes\sigma_{A'B'} = \sum q_i \sigma_{ABA'B'}$, then
$\rho_{AB}$ is decomposed as $\sum q_i \sigma_{AB}^{(i)}$ and
$\sigma_{A'B'} = \sum q_i \sigma_{A'B'}^{(i)}$.
We notice that even a smaller $\mathcal{M}$-decomposition for each
$\sigma_{AB}^{(i)}$ can be found provided that for  a generic  $\tilde\rho$,
\beq
\label{condfu}\mathcal{P}(\tilde\rho_{AB}) \geq \min \sum p_i
\mathcal{P}(\tilde\rho_{AB})^{(i)} \ .
\eneq
Also notice that this
minimization, however, cannot be possible unless constraints i) and ii) are
relaxed for this reduced density matrix.
In so doing, we require $\min \sum p_i \mathcal{P}(\rho_{s's})^{(i)} = 0$ for $s=A,B$,
thus releasing the constraints on
$\sigma_{AB}^{(i)}$ and $\sigma_{A'B'}^{(i)}$ and allowing the possibility of a
smaller minimum. At the same time we are obtaining the
absolute minimum, zero, for decompositions of $\rho_{s's}^{(i)}$ for
$s=A,B$, hence obtaining the minimizing decomposition
$\rho_{AB}\otimes\sigma_{A'B'} = \sum q_i \rho_{AB}^{(i)} \otimes
 q_j\sigma_{A'B'}^{(j)}$, which proves the statement.
We want to show that, given an arbitrary decomposition,
we can always build a decomposition with a smaller value of the
convex roof extension of $\mathcal{M}$:
\begin{eqnarray}
\nonumber \mathcal{M}(\sigma\otimes \eta) &=&\mathcal{A}  \min \sum_i p_i \sum_{mn} P(\sigma^{(i)}_{m n })\\
\nonumber &\geq &\mathcal{A} \min \sum_i p_i \left(\sum_{m'n'} P(\sigma^{(i)}_{m'n'}) +  \sum_{mn} P(\sigma^{(i)}_{m n })\right)\\ \nonumber
 &\geq & \mathcal{E}(\sigma\otimes 0_E) + \mathcal{E}(0_E \otimes \eta) \ ,
\end{eqnarray}
\vspace{-1.7cm}
\begin{equation}
\end{equation}
where $\mathcal{A} \equiv \mathcal{N} (C^N_2)^{-1}$.
For arbitrary
dimension a similar analysis holds, and Eq.~\eqref{condfu} now
reads,
\beq
\label{condfumul} \sum_{A,B} \mathcal{P}(\rho_{AB}) \geq \min \sum
p_i \sum_{A,B} \mathcal{P}(\rho_{AB})^{(i)} \ .
\eneq
Equation~\eqref{condfumul} is then a generalization of   Eq.~\eqref{condfu} for multipartite density matrices. Thus, the strategy provides a fully additive measure for
pure states. In addition, if both Eq.~\eqref{condfu} and Eq.
\eqref{condfumul} hold we also have  full additivity for mixed
bipartite and arbitrary mixed states respectively.

Our construction also guarantees strong super additivity. Note that
only a few of the measures reported in the literature satisfy both
the properties of additivity and strong super
additivity~\cite{Christandl}. For the
latter property to be satisfied we should have, for all
$\rho^{AA'BB'}$,
\beq E(\rho^{AA'BB'}) \geq E(\rho^{AB}\otimes 0_E^{A'B'}) +
E(0_E^{AB} \otimes \rho^{A'B'}) \ ,
\label{sa}
\eneq
where $\rho^{AB} = \Tr_{non(AB)}
\rho^{AA'BB'}$~\cite{Christandl}, that is tracing all
subsystems but $A,B$.
We extend this definition to the multipartite case, allowing
$A,B,A'$ and $B'$ to be multi-qubit registers. The proof that our
measure satisfies Eq.~(\ref{sa}) is as follows: Let $A,A',B$ and
$B'$ be the qubit registers $\{i_A\},\{i_{A'}\},\{i_B\}$, and
$\{i_{B'}\}$ respectively, and let $N_S = {\rm dim} \{i_S\}$, with
$S =A,A',B,B'$. Suppose that the decomposition minimizing
$\mathcal{M}(\rho^{AA'BB'})$ is $\rho^{AA'BB'}=\sum p_i
\sigma^{AA'BB'}_i$. Then
\begin{widetext}
\begin{eqnarray}
\mathcal{M}(\rho^{AA'BB'}) & = &  \mathcal{A} \Big(\sum
\mathcal{P}(i_A,j_{A})+ \sum \mathcal{P}(i_A,j_{A'}) + \sum
\mathcal{P}(i_A,j_{B})+   \sum \mathcal{P}(i_A,j_{B'}) + \sum
\mathcal{P}(i_{A'},j_{A'}) + \\ && \nonumber \sum
\mathcal{P}(i_{A'},j_{B})+\sum \mathcal{P}(i_{A'},j_{B'}) +   \sum
\mathcal{P}(i_{B},j_{B}) + \sum \mathcal{P}(i_{B},j_{B'}) + \sum
\mathcal{P}(i_{B'},j_{B'})\Big) \left(\rho^{AA'BB'}\right),
\end{eqnarray}
\end{widetext}
noting that $N\equiv N_A + N_B+N_{A'}+N_{B'}$, and that we require
that Eq.~\eqref{condfumul} holds. The case where
$N_A=N_B=N_{A'}=N_{B'}= 1$ is of particular interest, and the proof
requires Eq.~\eqref{condfu} to be satisfied.

 On the other hand,
analogously for $AB$ and $A'B'$,
\begin{eqnarray}
\mathcal{M}\left(\rho^{AB} =  \sum p_i \sigma_i^{AB}\right)  \leq  \mathcal{A} \Big(\sum \mathcal{P}(i_A,j_{A})+  \nonumber  \\
  \sum \mathcal{P}(i_A,j_{B})+ \sum
\mathcal{P}(i_{B},j_{B})\Big) \left(\rho^{AA'BB'}\right)\ .
\end{eqnarray}
This is so because, although we had already chosen a decomposition
minimizing $\mathcal{M}(\rho^{AA'BB'})$,  $\sum p_i \sigma_i^{AB}$ may not be the minimizing decomposition of
$\rho^{AB}$. Then,
\begin{eqnarray}
\mathcal{M}\left(\rho^{AA'BB'}\right) \geq  \mathcal{A}
\Big(\sum \mathcal{P}(i_A,j_{A})+ \sum \mathcal{P}(i_A,j_{B}) +   \nonumber \\
\sum\mathcal{P}(i_{B},j_{B}) + \sum \mathcal{P}(i_{A'},j_{A'}) +
\sum
\mathcal{P}(i_{A'},j_{B'}) + \, \, \, \, \,  \, \, \, \,  \, \nonumber \\
\sum \mathcal{P}(i_{B'},j_{B'})\Big) \left(\rho^{AA'BB'}\right)
  \geq   \mathcal{M}(\rho^{AB}) + \mathcal{M}(\rho^{A'B'}).  \, \, \, \, \,
\end{eqnarray}
The probe quantity must satisfy Eq.~\eqref{condfumul}
and Eq.~\eqref{condfu} to be strongly super additive in the most
general way, i.e., $\mathcal{M}(\rho^{AA'BB'CC'}) \geq
\mathcal{M}(\rho^{ABC}\otimes 0_E^{A'B'C'}) + \mathcal{M}(0_E^{ABC}
\otimes \rho^{A'B'C'}$). If it only satisfies Eq.~\eqref{condfu}
then we would have a restricted strong super additivity: a pairwise
strong super additivity, i.e., $\mathcal{M}(\rho^{A B C A' B' C'})
\geq \mathcal{M}(\rho^{AA'}\otimes 0_E^{BB'CC'}) +
\mathcal{M}(0_E^{AA'} \otimes \rho^{BB'} \otimes 0_E^{CC'}) +
\mathcal{M}(0_E^{AA'BB'}\otimes \rho^{CC'} )$. We show that one of
our proposed quantities satisfies Eq.~\eqref{condfu},  making it
a suitable candidate for our measure.

Other properties, such as regularizability, continuity, and non
lockability~\cite{Christandl}, may also be easier to prove through
our strategy. Thus, we have proposed a consistent and complete
measure of entanglement. The way that the additivity and the strong
super additivity introduced here differ from the traditional
presentation can be settled by replacing the
$\mathcal{M}(\ket{GHZ_N}) = \log_2d$ with $\mathcal{M}(\ket{GHZ_N})
= C^N_2 \log_2d$, and thus removing the $\mathcal{N}$ and
$(C^N_2)^{-1}$ normalization factors in $\mathcal{M}$; this would
leave us with a fully additive and strongly super additive measure
in the traditional way. We next discuss some possible choices for
$\mathcal{P}$.

{\it Quantifying entanglement via probe quantities
$\mathcal{P}$}---. We are interested in a quantity that, unlike the
tangle (which would yield zero for the two-qubit density matrices of
a generalized GHZ state, thus not being a desirable quantity to
average), is capable of determining how far a density matrix is from
being written as $\rho_{AB} = \rho_A \otimes \rho_B$.
Note that we require that it vanishes if and only if the matrix is factorizable,
i.e. if the two subsystems are not correlated. In contrast,
the concurrence vanishes if and only if the two subsystems are not
quantum correlated. Consider for example a GHZ and a fully
factorizable state: for the GHZ state, all two qubit density matrices
have vanishing quantum correlations whilst they have non-vanishing
values for $\mathcal{P}$; the fully separable state has
vanishing quantum correlations and vanishing $\mathcal{P}$. This
illustrates why $\mathcal{P}$ and not the concurrence suits our
strategy better.
We next
propose  two  quantities that satisfy the above requirements.

i) {\it The quasi-concurrence $\mathcal{Q}_C$}.
Following Wooters~\cite{Wooters-tangle},
consider, in decreasing order,  the eigenvalues $\lambda_i$'s of the
matrix $\sqrt{\rho_{AB} \tilde\rho_{AB}}$, where $\tilde\rho_{AB} =
(\sigma_2\otimes\sigma_2) \rho_{AB}^* (\sigma_2\otimes\sigma_2)$.
The concurrence is defined through the  $\lambda_i$'s  as
$C(\rho_{AB})= \max \{0,\lambda_1 -\lambda_2-\lambda_3-\lambda_4\}$,
and it can be shown to be equivalent to $2(1-\Tr [\rho_A]) =
\lambda_1$ for
 the pure state case. For the  above mentioned reasons,
 we define an alternate nonnegative
quantity, the {\it quasi-concurrence}
$\mathcal{Q}_C(\rho_{AB})= \lambda_1 + \lambda_2 -
\lambda_3-\lambda_4$. It is easy to convince oneself that it equals
zero for any factorizable density matrix, which follows from the
observation that if a matrix is factorizable then $\lambda_1
=\lambda_2=\lambda_3=\lambda_4$, thus yielding
$\mathcal{Q}_C(\rho_{AB}) = 0$. It equals one if and only if
$\sqrt{\rho_{AB} \tilde\rho_{AB}}$ has at most two non-vanishing
eigenvalues summing one, a condition satisfied by  EPR density
matrices and the two-qubit reduced density matrices in a generalized
GHZ state. As the eigenvalues are invariant under local unitary
operations, then $\mathcal{Q}_C(\rho_{AB})$ is LU invariant. It
also satisfies $\mathcal{Q}_C(\rho_{AB}) =
\lambda_1 + \lambda_2 -\lambda_3 - \lambda_4 \geq
\lambda_1-\lambda_2 -\lambda_3-\lambda_4 = \min \sum_i p_i
C(\rho^i_AB) = \min \sum_i p_i \mathcal{Q}_C(\rho^i_AB)$, as
$\rho^i_{AB}$ are pure density matrices, and thus $C(\rho^i_{AB})
=\mathcal{Q}_C(\rho^i_{AB})$. Hence $\mathcal{Q}_C(\rho_{AB})$
satisfies Eq.~\eqref{condfu}. Therefore a measure based on the
strategy proposed here using $\mathcal{Q}_C(\rho_{AB})$ is both additive
and pairwise strongly super additive~\footnote{We conjecture that
$\mathcal{Q}_C(\rho_{AB})$ also satisfies Eq.~\eqref{condfumul},
hence is both fully additive and strongly super additive.}.

ii) {\it The von Neumann's mutual information $\mathcal{F}r(A,B)$}.
This measures how correlated the two subsystems in $\rho_{AB}$ are;
if $S(\rho)$ denotes the von Neumann entropy then
$\mathcal{F}r(A,B)= \frac{1}{2}
\left[S(\rho_{A})+S(\rho_{B})-S(\rho_{AB})\right]$ ~\cite{Cerf
Adami-mutual information}. In this way, it is 0 if and only if
$\rho$ is factorizable and 1 for maximally non-factorizable density
matrices. Then $\mathcal{M}_{\mathcal{F}_r}$ is a local unitary
invariant  which measures how much information  we gain on average
after measuring one qubit, thus a suitable measure. It is an LOCC
non-increasing function for $N \leq 3$; numerical results suggest it
is also an LOCC monotone for $N>3$, however a formal proof is yet to
be provided. Also it can be seen it satisfies Eq.~$\eqref{condfu}$,
as $\mathcal{F}r$ measures total correlations which are greater or
equal than the quantum correlations obtained through the convex-roof
construction. We conjecture it also satisfies Eq.~$\eqref{condfumul}$
which allows us to build a fully additive measure
$\mathcal{M}_{\mathcal{F}r}$, with the nice feature that it can be
readily applied to qudit systems.

It is not yet fully understood which of the above given measures
more accurately quantifies the degree of factorizability, however,
and for illustrative purposes, we perform below some numerical
calculations using $\mathcal{Q}_C(\rho_{AB})$ and
$\mathcal{F}r(A,B)$ without necessarily implying that one of them is
the most accurate quantity. We have performed numerical simulations
which suggest that $\mathcal{F}r(A,B)$ and $\mathcal{Q}_C
(\rho_{AB})$ make $\mathcal{M}$ an LOCC non-increasing function.

First, we use $\mathcal{Q}_C(\rho_{AB})$  as the probe quantity. Let
us consider the case of multipartite pure qubit states. i) For
generalized GHZ states,
$\ket{GHZ} = \frac{1}{\sqrt{2}}\big(\ket{0}^{\otimes
N}+\ket{1}^{\otimes
N}\big)$ (consider $N \geq 3$), we have
$\mathcal{Q}_C(\rho_{AB}) = 1$ for all $\rho_{AB}$, thus yielding
$\mathcal{M} = 1$, ii) for  a tripartite W state, $\ket{W} = \frac{1}{
\sqrt{3}} \big(\ket{100}+\ket{010}+\ket{001}\big)$,  $\mathcal{M} =
2/3$, and iii)  for a completely separable state we have
$\mathcal{Q}_C(\rho_{AB}) = 0$ for all $\rho_{AB}$.
Second, we test for the $\mathcal{F}r$'s. Considering the
same states as above, we get i) all the two qubit reduced density
matrices yield $\mathcal{F}r(A,B) = 1/2$, ii) $\mathcal{F}r(A,B)\sim
0.46$ for all $(A,B)$, and iii) $\mathcal{F}r(A,B) =0$,
respectively. Here the result of case i) may seem a non-desired one,
as there would be states with $\mathcal{F}r(A,B)$'s greater than 1/2.
 It is interesting to note, however, that
this observation leads us to the following finding. Consider the maximally entangled mixed state
(MEMS)~\cite{MEMS}:
\beq \rho_{MEMS}=\left(
                                                             \begin{array}{cccc}
                                                               x/2 & 0 & 0 & x/2 \\
                                                               0 & 1-x & 0 & 0 \\
                                                               0 & 0 & 0 & 0 \\
                                                               x/2 & 0 & 0 & x/2 \\
                                                             \end{array}
                                                           \right) \ .
\eneq
A straightforward calculation shows that $\mathcal{F}r(A,B)
> \mathcal{F}r(A,B)_{\ket{GHZ}}$ for $x \rightarrow 1$. This does
not imply, however, that we would have a value of $\mathcal{M}$
higher than that of $\mathcal{M}_{\ket{GHZ}}$. To see this, consider
the smallest purification of $\rho_{MEMS}$~\cite{Nielsen-Book},
given by the four qubit state $\ket{\Psi}_{pure MEMS} = \sqrt{1-x}
\ket{0101} + \sqrt{x/4} (\ket{0000} + \ket{0011} + \ket{1100} +
\ket{1111})$. Clearly $\mathcal{F}r(1,2) = \mathcal{F}r(3,4) >
\mathcal{F}r(A,B)_{\ket{GHZ}}$ for $x\rightarrow 1$, but
$\mathcal{F}r(1,3) = \mathcal{F}r(1,4) = \mathcal{F}r(2,3) =
\mathcal{F}r(2,4) < \mathcal{F}r(A,B)_{\ket{GHZ}}$, and $\mathcal{M}
< \mathcal{M}_{\ket{GHZ}}$, thus illustrating the point. Here we
have fixed the normalization constants in such a way that
$\mathcal{F}r(A,B)_{\ket{EPR}} =1$. The following theorem formalizes
the above observation

\begin{theorem}
\label{normal} For a N-qudits quantum state,
$\mathcal{M}_{\mathcal{F}r}$ is normalized to $\log_2d$.
\end{theorem}
\begin{proof}
For $\mathcal{F}r$, we can prove that the measure is indeed
normalized in the following way. To simplify the notation, we shall
use $S(X) \rightarrow X$. Von Neumann entropy's strong sub
additivity reads \beq XYZ \leq XY + YZ - Y, \eneq we will use this
inequality intensively using different partitions at our convenience
along the rest of the paper. The proof for the three qubit case
$\Psi_{ABC}$ is trivial, using  S(AB) = S(C), we get that
$\mathcal{M} = \mathcal{N} (C^3_2)^{-1} (A+B+C) \leq \log_2 d$. For
the four qubit case, using von Neumman's entropy strong
subadditivity and assignations of $(X,Y,Z) =
\{(B,A,C);(B,D,A);(B,C,D)\}$ it follows that
\begin{eqnarray}
\nonumber \mathcal{M}_{\mathcal{F}r} &\leq& \frac{2}{12} (3 B+ 2A +2C+ 2D-BAC - BDA - BCD)\\
\nonumber &=& \frac{1}{6} (3 B + A + C+ D) \leq \log_2d
\end{eqnarray}
where we have used that $S(\rho_i) \leq \log_2 d$.

 For five qubits, consider the following
inequality,
\begin{eqnarray}
\nonumber XYZW &\leq& YXZ + YWZ -YZ\\
\nonumber &\leq & XY + XZ + YW + WZ - YZ - X - W,
\end{eqnarray}
summing for the assignations of $(X,Y,Z,W) = \{(E, A, B, C);(E, A,
C, D);(B, A, D, C);(B, A, E, D);$ $(A, B, C, D);(A, B, D, E);(D, B,
E, C);(B, C, D, E);$ $(A, C, E, B);(A, D, E, C) \}$ we get, using
that in a N-qubit pure state $S(A_1,...,A_m) = S(A_{m+1},...,A_N)$

\begin{widetext}
\beq
\nonumber  6(A +B+C+D+E) - 3(AB + AC + AD+AE+BC+BD+BE+CD+CE+DE)\leq 0\\
\eneq \beq
\nonumber 12(A+B+C+D+E)-3(AB+AC+AD+AE+BC+BD+BE+CD+CE+DE)\leq 6(A +B+C+D+E)\\
\eneq\beq \mathcal{M}_{\mathcal{F}r} \leq \log_2 d. \eneq
\end{widetext} For higher number of qubits, similar inequalities can
be tailored.\qed
\end{proof}

Hence, we see that according to $\mathcal{F}r(A,B)$, a GHZ state is
not the state with the highest non-factorizability among its
components but the one with the highest average, and thus with the
highest possible $\mathcal{M}$. Moreover, if we use
$\mathcal{Q}_C(\rho_{AB})$ as the probe quantity, we obtain a
stronger condition: here the GHZ state not only has the highest
average but the highest non-factorizability among its components.

Hence, we identify two types of states. i) {\it Homogeneously}
entangled states: the ones for which $\mathcal{P}(A,B)$ is the same
for all $(A,B)$, and ii) {\it Heterogeneously} entangled states: the
ones for which the values for $\mathcal{P}(A,B)$ may be different
for each $(A,B)$. In the first category we find, e.g., the W state,
and the GHZ state, whilst in the second one we find $m$-separable
states, $\ket{EPR} \otimes \ket{EPR}$, $\ket{\Psi}_{pure MEMS}$,
etc. In this way, we introduce the following definition:  a GHZ
state is the homogeneously entangled state with the highest value of
$\mathcal{F}r(A,B)$ $(\mathcal{Q}_C(\rho_{AB}))$ and thus  of
$\mathcal{M}$.

We next compute for some states that are directly relevant to quantum
information protocols~\cite{Bennett}. We consider the ``$(1,4)\otimes(2,3)$" correlated product of EPR states $\ket{\Psi} =
\frac{1}{2} (\ket{0000} + \ket{0110} + \ket{1001} +
\ket{1111})$~\cite{Lee-twoteleportation,Rigolin-twoteleportation,YeoChua-twoteleportation},  the  four-qubit entangled state
$\ket{\chi} = \frac{1}{2 \sqrt{2}}( \ket{0000} -
\ket{0011}-\ket{0101}+\ket{0110}+\ket{1001}+\ket{1010}+\ket{1100}+\ket{1111})$~\cite{YeoChua-twoteleportation},
 and the so-called cluster state $ \ket{\Phi_4} = \frac{1}{2} (\ket{0000} +
\ket{0110}+\ket{1001}-\ket{1111})$~\cite{cluster-states}. A direct
calculation yields\\

\hspace{-1.2cm}
\begin{tabular}{c c c c c c c c c}
  \hline   \hline
 $\mathcal{Q}_C [\mathcal{F}r]{(i,j)}$ &$\small{(1,2)}$&$(1,3)$&$(1,4)$&$(2,3)$&$(2,4)$&$(3,4)$&$   \mathcal{M}_{\mathcal{Q}_C}$&$\mathcal{M}_{\mathcal{F}r}$\\ \hline
  $\ket{\Psi}$ & 0 & 0 & 1 [1] & 1 [1]  & 0 & 0 & 1/3 & 2/3 \\
  $\ket{\chi}$ & 0 & 0 & 1 [1/2]  & 1 [1/2]  & 0 & 0 &  1/3 & 1/3\\
  $\ket{\Phi_4}$ & 0 & 0 & 1 [1/2]  & 1 [1/2]  & 0 & 0 & 1/3 & 1/3\\
  \hline  \hline 
\end{tabular}
\vspace{0.1cm}\\

\noindent which shows that $\mathcal{Q}_C(A,B) \neq 0$, and
$\mathcal{F}r(A,B) \neq 0 $ for $(A,B) = (1,4),(2,3)$, and equal
zero otherwise. Thus,
there are non-factorizable states {\rm (e.g. $\ket{\chi}$ and
$\ket{\Phi_4}$)} whose probe quantities yield a similar  structure to that of the
semi-factorizable states {\rm (e.g. $\ket{\Psi}$)}. Equivalently,
we may define a {\it genuine globally} entangled state as a
state for which $\mathcal{P}(A,B)\neq 0$ for all $(A,B)$. Note that
$\mathcal{F}r$ gives  the mutual information of system AB, i.e.
how much information we gain about B after measuring A or vice versa.
Our definition would read  that a state
is genuine globally entangled if after measuring one qubit we gain
some information about all of the other qubits in the register. In the case of
the four-party states considered here, we see that we gain no
knowledge of the state of qubits 2 or 3 after measuring qubits 1 or
4, i.e. qubits 2 and 3 are not correlated with 1 and 4 (see table above), and they
behave, in this context, as if they were semi-factorizable and not like
genuinely globally entangled states. This definition is in contrast with results reported
in the literature \cite{YeoChua-twoteleportation,Oliveira-GEM}, in
 particular with bipartition strategies such as the ones introduced in Refs.
\cite{Scott,Oliveira-GEM}.

{\it Remarks on the additivity of  $\mathcal{M}$}---.
We note that the way we have casted additivity is not equivalent to
the traditional way
\beq \label{tradadd} E (\rho\otimes \sigma) = E
(\rho) + E (\sigma) \ .
\eneq
The choice we have made here is based on the
condition $E(\ket{GHZ_N}) = \log_2 d$, which implied the inclusion
of $\mathcal{N}$ and $(C^N_2)^{-1}$ as normalization factors.
However, by choosing the normalization
\beq E(\ket{GHZ_N}) =  C^N_2 \log_2 d, \eneq
we
remove the normalization factors and hence have a fully additive measure
in the traditional way. This observation is of relevance, and the same
idea of the previously presented proofs holds by means of
omitting the normalization factors and thus obtaining
Eq.~\eqref{tradadd}.

We want to argue in favor of the way we introduced additivity in the
first part of this work. Additivity is a desired property based on the intuitive
observation that if Alice and Bob have two EPR pairs they must have
twice the amount of entanglement than if they only share one EPR pair. Our formulation
 is consistent with this observation, but an
important consideration regarding the size of the qubit register holds.
In order to compare two EPR pairs with one EPR
pair we must consider the right scenario, and the question can be posed
in the following terms. Suppose Alice and Bob share an EPR pair, hence they
have a fixed amount of entanglement. If they manage to create
another EPR pair, then how much entanglement do they share now? The answer is, as we have shown earlier,
that they should double the amount of entanglement.
However, if they can create a second EPR pair, it means that they
had at least four qubits at their disposal in the first place, which
has been illustrated by the $\otimes 0_E$ and $0_E\otimes$ terms in our formula.
Thus, our formulation of additivity indeed shows  that the strategy provides
an additive measure of entanglement. For completeness we cast the
traditionally additive measure we reach after the discussion above
\beq \label{trad} \mathcal{M}^T_{\mathcal{P}} = \sum_{(A,B)}
\mathcal{P} (A,B) \ .
\eneq
Either of the two measures, $\mathcal{M}$ or $\mathcal{M}^T$,
depends on the choice of additivity we make. For illustration we now
recast the proof of additivity in the traditional way. Recall that
we have already shown that $\mathcal{F}r$ and $\mathcal{Q}c$ satisfy
Eq.\eqref{condfu}, and thus we have a fully additive measure.

\begin{theorem}
(Traditional additivity)\\
The entanglement measure $\mathcal{M}^T_\mathcal{P}$ is fully
additive, i.e. \beq \label{add} \mathcal{E}(\sigma \otimes \eta) =
\mathcal{E}(\sigma) + \mathcal{E}( \eta) \ , \eneq  provided \beq
\label{condfumul} \sum_{A,B} \mathcal{P}(\rho_{AB}) \geq \min \sum
p_i \sum_{A,B} \mathcal{P}(\rho_{AB})^{(i)} \ . \eneq
\end{theorem}

\begin{proof} The proof basically relies on the pairwise minimizing
condition. Consider two generic $m$ and $N-m$ qudit density
matrices, $\eta$ and $\sigma$, thus it guarantees that there exists
a bifactorizable decomposition of the form $\eta\otimes\sigma = p_i
\eta^i \otimes q_i \sigma^i$. Our plan or the proof is the
following: we will assume that we have a generic,
non-bifactorizable, minimizing decompostion, and we will show that
the bifactorizable decomposition has a lower value of entanglement
following the convex roof construction recipe. For simplicity we
will show it here for the $m=2$ case, however the argument is easily
extrapolated to the multipartite case.

If we have a non-bifactorizable decomposition of the form $\rho =
\sum p_i \sigma^i$, then we have that there are non-vanishing
pairwise minimizing decompositions for $\rho_{12}$,
$\rho_{13}$,$\rho_{14}$,$\rho_{23}$,$\rho_{24}$ and $\rho_{34}$,
with values denoted as $f(\rho_{AB})$. On the other hand, a
bifactorizable decomposition would have other values for their
minimizing decomposition, namely $g(\rho_{AB})$, and in particular
some of them vanish,
$g(\rho_{13})=g(\rho_{14})=g(\rho_{23})=g(\rho_{24})=0$, which
implies that $g(\rho_{13}) \leq f(\rho_{13})$ and similarly for
$(1,4)$,$(2,3)$ and $(2,4)$.

Now, the proof would be complete if we demand that $g(\rho_{12})
\leq f(\rho_{12})$ and $g(\rho_{34}) \leq f(\rho_{34})$. This is
equivalent to demand that the lowest value achieved by any
decomposition is on pure state matrices, namely on a decomposition
$\rho_{AB} = \sum_\alpha p_\alpha \rho_{AB}^\alpha$ where
$\rho_{AB}^\alpha$ is a pure density matrix. This is equivalent then
to demand that
\beq \label{condfu}\mathcal{P}(\tilde\rho_{AB}) \geq \min \sum p_i
\mathcal{P}(\tilde\rho_{AB})^{(i)} \ . \eneq
as claimed. The extension of the argument to more qudits is
straightforward, and would leave us with the condition \beq
 \sum_{A,B} \mathcal{P}(\rho_{AB}) \geq \min \sum
p_i \sum_{A,B} \mathcal{P}(\rho_{AB})^{(i)} , \eneq where the
minimization is intended over every possible decomposition on pure
states. Note that if this is true then any decomposition on mixed
states will yield a higher value of entanglement.\qed
\end{proof}

We have presented an alternative approach for quantifying
multipartite entanglement. In so doing, we have proposed
entanglement measures based on pairwise strategy which naturally
exhibit, in addition to the usual entanglement properties
~\cite{Plenio-entcond,Vedral-entropy,Christandl}, additivity and
strong super additivity. We have given a set of measures which
fulfill the role of $\mathcal{P}(A,B)$ and are LOCC non-increasing
functions according to numerical results. Our proposal shows that
the pairwise analysis strategy is very effective for quantifying
entanglement and that it also guarantees most of the
non-conventional properties of entanglement measures. In addition,
we have also shown that such a strategy allows a redefinition of the
structural concept of global entanglement.

We thank T. Brandes and C. Emary for useful comments on the manuscript.
We gratefully acknowledge financial support from COLCIENCIAS under
Research Grants No.~1106-14-17903 and No.~1106-05-13828. GAPS thanks
G. H. Paz, I. Silva, G. R. Paz and D. F. Gutierrez for support.

\end{document}